\documentclass{emulateapj}

\usepackage{apjfonts}

\slugcomment{Accepted for ApJ}

\newcommand{\bicho}{1FGL J1625.8$-$2429c}

\newcommand{\grp}    {${\rlap.}^{\circ}$}

\newcommand{\ltsima} {$\; \buildrel < \over \sim \;$}
\newcommand{\simlt}  {\lower.5ex\hbox{\ltsima}}            
\newcommand{\gtsima} {$\; \buildrel > \over \sim \;$}
\newcommand{\simgt}  {\lower.5ex\hbox{\gtsima}}            

\shorttitle{Are T Tauri stars gamma-ray emitters?}
\shortauthors{del Valle et al.}

\begin{document}
\title{Are T Tauri stars gamma-ray emitters?}


\author{Mar\'{\i}a Victoria del Valle\altaffilmark{1,2}, Gustavo E. Romero\altaffilmark{1,2}, Pedro Luis Luque-Escamilla\altaffilmark{3}, Josep Mart\'{\i}\altaffilmark{4} and Juan Ram\'on S\'anchez-Sutil\altaffilmark{4}}

\altaffiltext{1}{Instituto Argentino de Radioastronom\'{\i}a (IAR), CCT La Plata  (CONICET), C.C.5, (1894) Villa Elisa, Buenos Aires, Argentina.}
\altaffiltext{2}{Facultad de Ciencias Astron\'omicas y Geof\'{\i}sicas, Universidad Nacional de La Plata, Paseo del Bosque s/n, 1900, La Plata, Argentina.}
\altaffiltext{3}{Departamento de Ingenier\'{\i}a Mec\'anica y Minera, Escuela Polit\'ecnica Superior de Ja\'en, Universidad de Ja\'en, Campus Las 
Lagunillas s/n, A3, 23071 Ja\'en, Spain.}
\altaffiltext{4}{Departamento de F\'{\i}sica, Escuela Polit\'ecnica Superior de Ja\'en, Universidad de Ja\'en, Campus Las Lagunillas s/n, A3, 23071 Ja\'en, Spain.}

\begin{abstract}
T Tauri stars are young, low mass, pre-main sequence stars surrounded by an accretion disk. These objects present strong magnetic activity and powerful magnetic reconnection events. Strong shocks  are likely associated with fast reconnection in the stellar magnetosphere. Such shocks can accelerate particles up to relativistic energies.
We aim at developing a simple model to calculate  the radiation produced by non-thermal relativistic particles in the environment of T Tauri stars. We want to establish whether this emission is detectable at high energies with the available or forthcoming $\gamma$-ray telescopes.
We assume that particles (protons and electrons) pre-accelerated in reconnection events are accelerated at shocks through Fermi mechanism and  we study the high-energy emission produced by the dominant radiative processes.
We calculate the spectral energy distribution of T Tauri stars up to high-energies and we compare the integrated flux obtained with that from a specific {\it Fermi} source, \bicho, that we tentatively associate with this kind of young stellar objects (YSOs).
We suggest that under reasonable general conditions nearby T Tauri stars might be detected at high energies and be responsible for some  unidentified {\it Fermi} sources on the Galactic plane.

\end{abstract}


\section{Introduction}
\label{s_intro}
The \textit{Fermi Gamma-Ray Space Telescope} was launched on 2008 June 11 and is the successor of the \textit{Compton Gamma Ray Observatory}. It has two $\gamma$-ray instruments on board, the Large Area Telescope (LAT) and the Gamma-ray Burst Monitor (GBM). The first one, a pair-production telescope, is the successor of the Energetic Gamma Ray Experiment Telescope (EGRET). LAT has a sensitivity $\sim$ 10$^{-9}$ ph s$^{-1}$ cm$^{-2}$ in the energy range from 20 MeV to over 300 GeV.  

Based on the first eleven months of survey, the {\it Fermi} First Year Catalog has been published (Abdo et al. 2010). The catalog consists of 1451 sources  with  statistical significances of 4${\sigma}$ or higher. This catalog provides potential identifications, when available.  
Nearly 630 {\it Fermi} sources remained unidentified (Abdo et al. 2010). A fraction of these unidentified sources lay on the Galactic plane (latitudes $\leq$ $5^{\circ}$). They may be pulsars, supernova remnants (SNR), the effect of strong winds from massive stars, high-mass or low-mass X-ray binaries (HMXBs or LMXBs), or stellar clusters, among other possibilities (e.g. Romero, Benaglia, \& Torres 1999). 

Munar-Adrover, Paredes, \& Romero (2011) have crossed the {\it Fermi} First Year Catalog with catalogs of known YSOs, in order to identify those protostars that might emit $\gamma$-rays. They conclude that 72\% of the candidates obtained by spatial correlation should be $\gamma$-ray sources with a confidence above 5$\sigma$. Massive YSOs have already been  claimed to be $\gamma$-ray  sources (e.g. Araudo et al. 2007, Bosch-Ramon et al 2010). However, such a claim has not been done yet for low-mass protostars.

In this paper we shall argue that these stars, in particular T Tauri stars, can be faint but sometimes detectable $\gamma$-ray sources.
We will focus our attention on the physical processes that can generate $\gamma$-ray emission in T Tauri magnetospheres. Specifically, we shall discuss whether these protostars can be responsible for sources like 1FGL J1625.8$-$2429c. In the next section we present a brief introduction to T Tauri stars and we describe the general physical scenario. In Section 3 we provide the details of our  model.  In Section 4 we present our calculations  applied to a specific group of T Tauri stars and the main results. In Section 5,  we discuss these results and, finally, we close with some brief conclusions.  

\section{Scenario}
\label{jet-clump}

T Tauri stars are low-mass stars ($M$ $<$ 3 $M_{\odot}$) in their early stages of evolution. Classical T Tauri stars have typical K-M spectral types with $T_{\rm eff}$ $\sim$ 3000-5000 K (Montmerle \& Andr\'e 1989). Values of the stellar radius are $\sim$ 2$-$3 $R_{\odot}$. 
As progenitors of solar-like stars, they are subject of intense study. They  are  usually optically visible and the youngest objects drive bipolar outflows. 
The protostars are born in the collapse of molecular clouds. The cloud material has high angular momentum and therefore a circumstellar accretion disk is formed (e.g. Bekwith et al. 1990). 
These accretion disks are truncated in the vicinity of a co-rotation radius where the magnetic pressure equals the gas pressure in the disk (see Fig. \ref{fig:magnetosphere}). Infrared observations allow to infer inner radii of 0.07$-$0.54 AU (Muzerolle et al. 2003), consistent with the co-rotation radius. UV emission has been used to estimate the mass accretion rate, yielding typical values of $\sim$ 10$^{-8}$ $M_{\odot}$ year$^{-1}$ (Gomez de Castro \& Lamzin 1999; Johns-Krull et al. 2000). The infall velocity can almost reach the free-fall speed. 
The existence of a large amount of dust in a flattened disk is confirmed by intense emission from the micrometer to the millimeter bands. The disks can now be directly imaged (e.g. Dutrey, Guilloteau, \& Simon 1994). 
In addition, these objects drive strong winds with mass loss rate $\dot{M}$ $\sim 10^{-8}\, M_{\odot}$ year$^{-1}$ and velocity  $v_{\rm w}$ $\sim$ 200 km s$^{-1}$ (Feigelson \& Montmerle 1999).

X-ray studies indicate particle number densities of the accreting plasma of about $10^{12}$ cm$^{-3}$ (G\"{u}nther et al. 2007).  

Variable thermal X-ray emission is also detected from T Tauri stars in the keV band. 
Luminosities are found to be in the range $\sim$ 10$^{31}$ $-$ 10$^{33}$ erg s$^{-1}$.
This emission comes from a  high density plasma at a typical temperature of $\sim$ 10$^{7}$ K; flares with temperatures $\sim$ 10$^{8}$ K have  been detected (Tsuboi et al. 1998). These flares have durations of $\sim$ 10$^{3}$ $-$ 10$^{4}$ s. Such events are expected to occur in magnetic flux tubes with spatial extent of $\sim$ 10$^{10}$ $-$ 10$^{11}$ cm (e.g Hayashi et al. 1996).  
Models for the X-ray activity based on the interaction of the stellar object and the circumstellar disk  have been proposed by several authors (e.g.  Hayashi et al. 1996, Birk et al. 2000).

These X-rays flares are widely considered as upscaled versions of solar flares. 
The rapid heating and cooling of plasma and acceleration of particles must arise from efficient MHD processes, such as solar-type magnetic reconnection events in twisted flux tubes that connect the central object and the circumstellar disk (Birk et al. 2000).  
Magnetic reconnection is one of the fundamental processes in astrophysical plasmas because it explains large-scale, dynamic releases of magnetic energy. It is essentially a topological reconfiguration of the magnetic field caused by a change in the connectivity of the field lines. It is this change which allows the release of stored magnetic energy, which in many situations is the dominant source of free energy in a plasma.
Several  works have been done on particle acceleration through  magnetic reconnection ( e.g. Schopper, Lesch, \& Birk 1998, Zenitani \& Hoshino 2001, de Gouveia Dal Pino et al. 2010). Strong shocks  resulting from supersonic plasma ejections are the likely outcome of massive reconnection in T Tauri magnetospheres. Such shocks can in principle accelerate particles up to relativistic energies through Fermi mechanism.
  
The  expected  values of magnetic field in T Tauri  stars  are $\sim$ 1 kG (e.g. Johns-Krull 2007) and the field structure is complex and multipolar, as in the Sun. 
For simple flare models, quantitative properties of large-scale  magnetic field structures can be inferred (e.g. Montmerle et al. 1983, Walter \& Kuhi 1984). Assuming equipartition conditions, $B_{\rm eq}^{2}/8{\pi}$ = $2n_{\rm e}kT$, the magnetic field strength in the magnetosphere is approximately 10$^{2}$ G (Feigelson \& Montmerle 1999).

\section{Model}
\subsection{Particle acceleration}
 
Diffusive shock acceleration does not work for slow-mode shocks, consequently is thought that diffusive shock acceleration is not important for magnetic reconnection that involves slow-mode shocks (e.g. Priest \& Forbes 2000). Under some conditions, however,  any obstacle which obstructs the outflow will create a fast-mode shock (Tsuneta \& Naito 1998). In a T Tauri magnetosphere the obstacles might be clumps from the strong stellar wind; considerable observational evidence supports the idea that the wind structure is clumpy (e.g. Owocki \& Cohen 2006). Moreover, shock acceleration is now applied to the outflow regions of coronal
magnetic reconnection sites, where first-order Fermi mechanism at the standing fast
shock is a leading candidate (Aschwanden 2008). The shocks are expected to  accelerate charged particles  up to high energies by a Fermi-like diffusive process (e.g. Drury 1983). 

Additionally, recent extensive 3D numerical simulations performed by Kowal et al. (2011) show that within contracting magnetic islands or current sheets, charged particles are accelerated by a first-order Fermi process while outside the current sheets and islands the particles experience mostly drift acceleration due to magnetic field gradients. These results are supported by  observations of solar flares that suggest that magnetic reconnection should be first slow in order to ensure the accumulation of magnetic flux and then suddenly become fast to allow a rapid energy release (e.g. Lazarian \& Vishniac 1999).  Particles scattered by turbulence between converging magnetic mirrors formed by oppositely-directed magnetic fluxes moving towards each other at the fast reconnection speed can undergo diffusive acceleration without the requirement of strong shock formation (Kowal et al. 2011).  

Independently of the details of the acceleration mechanism, we assume that a population of non-thermal relativistic particles, electrons and protons, is injected into the magnetosphere of the T Tauri star. These particles will interact with the large scale magnetic fields, with the existing radiation fields, and with the magnetospheric plasma, producing non-thermal electromagnetic radiation. The size of the acceleration region is the spatial scale where reconnection takes place, i.e the flux tube length.

The power available in  the magnetized system is  

\begin{equation}
L = \frac{B^{2}}{8\pi}A \, \, v_{\rm A},
\end{equation}
where $A$  is the area of the flux tube, of length $l = 10^{11}$ cm and aspect ratio $\sim$ 0.1 $l$ (Feigelson \& Montmerle 1999), and $v_{\rm A}$ is the Alfven speed, $v_{\rm A} = \sqrt{{B^{2}}/({4\pi m_{p} n})}$, with $n$ the particle density and $m_{p}$ the proton mass.
For $B = 2\times 10^{2}$ G, we get $L$ $\sim$ $10^{34}$ erg s$^{-1}$. We assume that 10\%  of this power is released in the reconnection processes.  In turn, a fraction $q_{\rm rel}$ $\sim$ 10\% of this power goes to relativistic particles. These values are in accordance to  inferred values for the Sun, and can be considered even conservative. For instance, in large solar flares the accelerated particles contain up to 10-50\% or more of the total energy released, whereas in gradual events $\sim$ 10\% of the total power goes to the accelerated particles (see Lin 2008 and references therein).

The efficiency of non-thermal acceleration in the magnetized plasma is
\begin{equation} 
\eta \sim 10^{-1} \frac{r_{\rm g} c}{D}\left(\frac{v_{\rm rec}}{c}\right)^{2},
\end{equation}
mimicking the efficiency for standard first order Fermi acceleration theory behind shocks\footnote{This assumption holds only if there is a first-order Fermi mechanism occurring within the reconnection zone caused by two converging magnetic fluxes of opposite polarity which move to each other with $v_{\rm rec}$. This mechanism of  first-order Fermi acceleration occurring within a reconnection site was first proposed and described analytically in de Gouveia Dal Pino \& Lazarian (2005) and successfully tested through 3D MHD simulations by Kowal, de Gouveia Dal Pino, \& Lazarian (2011).} (Drury 1983,  Vila \& Aharonian 2009); here $D$ is the particle difussion coefficient, $r_{\rm g}$ is the particle gyroradius and $v_{\rm rec}$ is the reconnection speed. If $D$ is in the Bohm limit $D_{\rm Bohm}$ $=$ $r_{\rm g}c/3$. The  reconnection speed in violent reconnection events satisfy $v_{\rm rec}$ $\sim $ $v_{\rm A}$ (Lazarian \& Vishniac 1999, Kowal et al. 2009), so we assume $v_{\rm rec}$ $=$ $0.6v_{\rm A}$, that gives an efficiency  $\eta\sim10^{-6}$. This efficiency, although not very high, will allow maximum energies well into the $\gamma$-ray domain, as we will see.

\subsection{Energy losses}

The relativistic electrons and protons lose energy through different interactions. The maximum energy that they can reach depends on the processes of energy loss, the available space for effective acceleration and on the acceleration rate. This rate is $t_{\rm acc}^{-1} = \eta ecB/E$ for first order Fermi mechanism.

The main loss mechanism for electrons in the magnetosphere of T Tauri stars is synchrotron radiation. Other relevant loss processes are relativistic Bremsstrahlung ---the leptons interact with ions of material in the magnetosphere---, and inverse Compton scattering  (IC) --- electrons scatter up the ambient photons---. The main ambient photon field is the thermal X-ray emission, which corresponds to a temperature of $\sim$ $10^{7}$ K, and has a typical luminosity (in quiescence) of $\sim$ 6$\times 10^{29}$ erg s$^{-1}$ according to the observed values (see Table \ref{tableIII}). 

\begin{figure}
\includegraphics[angle=270, scale=0.5]{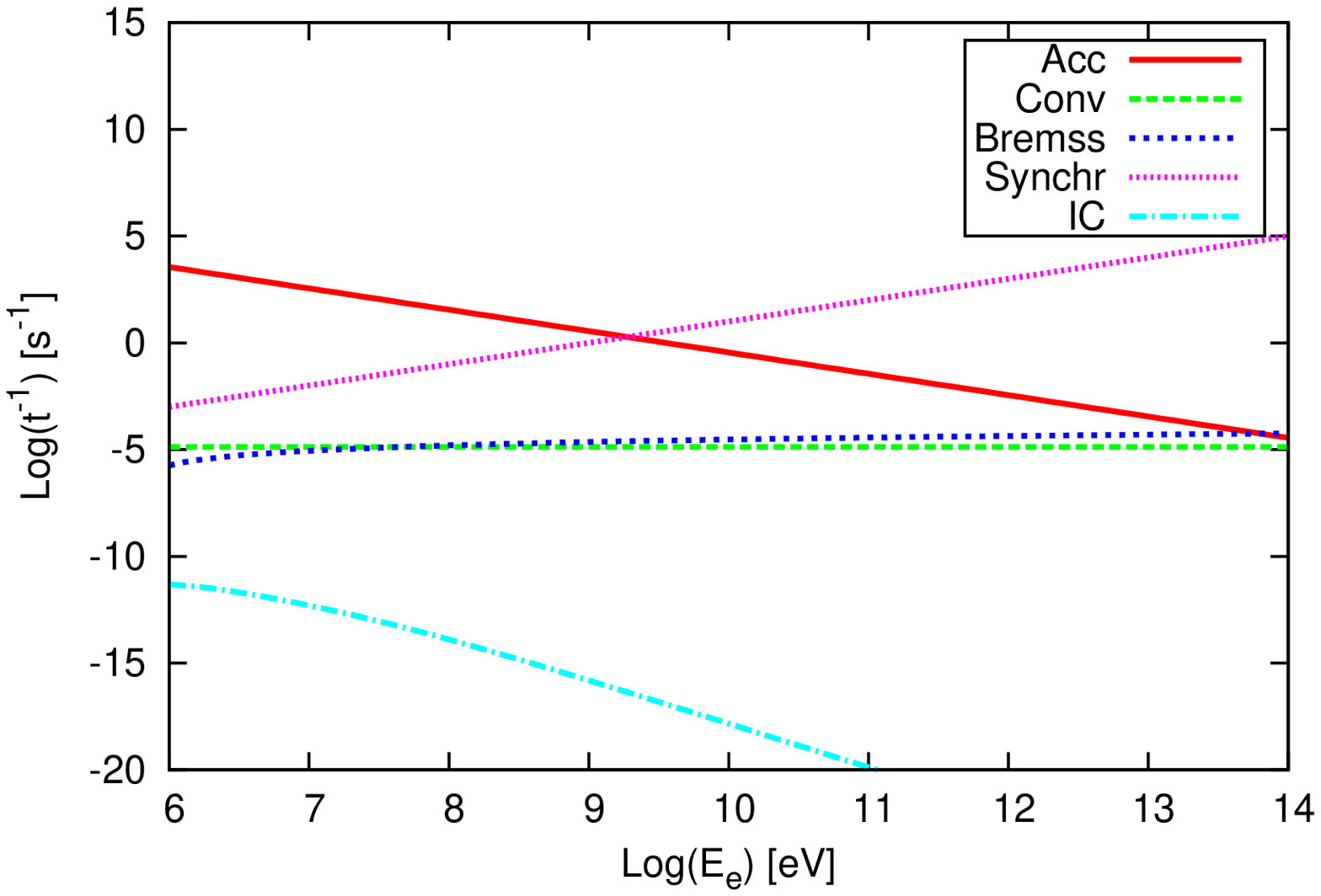}
\caption{Acceleration and cooling rates for electrons. The identification of the different curves is given in the box in the upper right corner.}
\label{fig:perdEle}
\includegraphics[angle=270, scale=0.5]{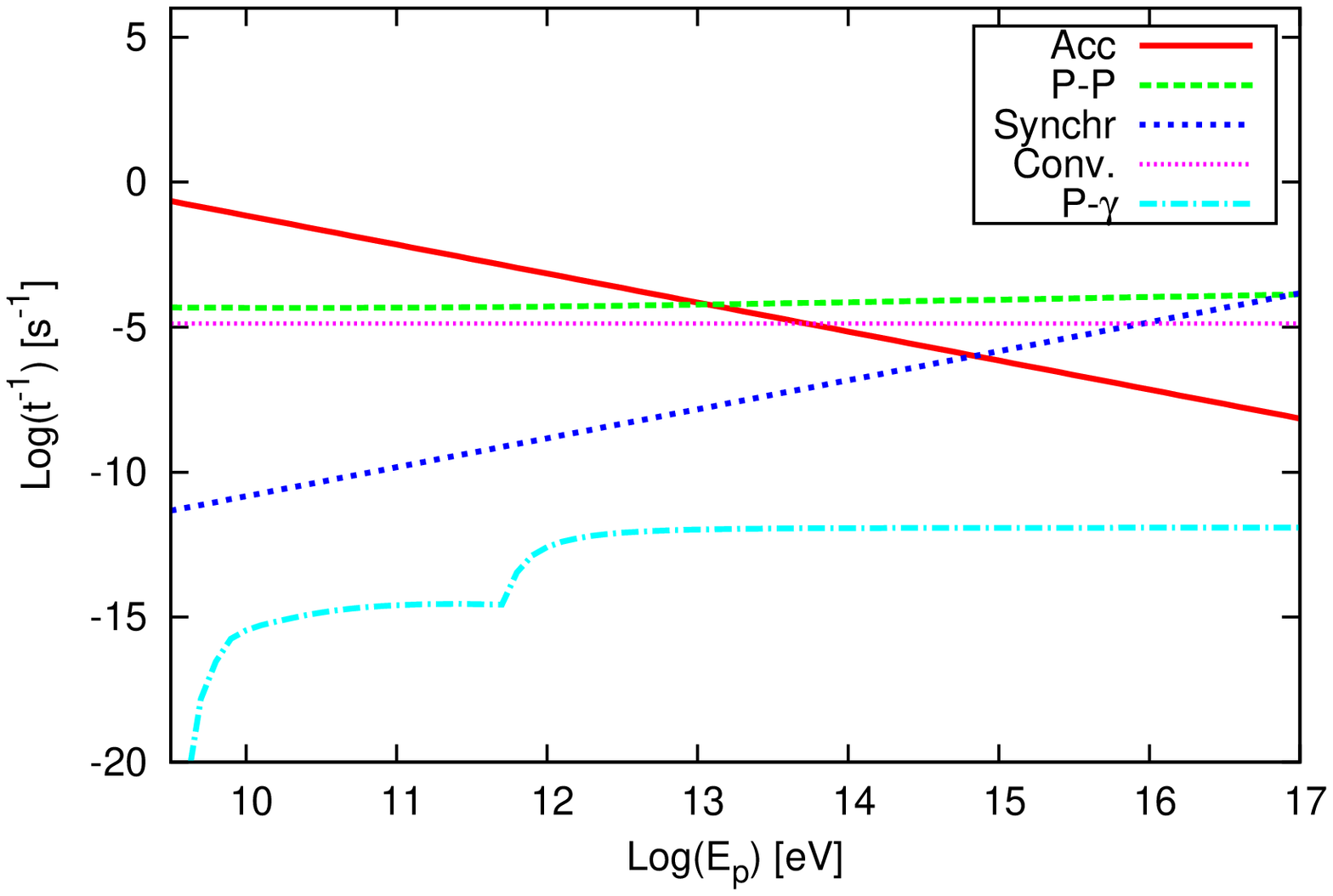}
\caption{Acceleration and cooling rates for protons.}
  \label{fig:perdPro}
\end{figure}

Figure  \ref{fig:perdEle} shows the cooling rates and the acceleration rate for electrons in the case of the parameters shown in Table 1. These parameters are fixed in accordance to the typical values discussed above for  T Tauri stars (see e.g. Feigelson \& Montmerle 1999). Particles can also escape by convection from the acceleration region with a rate $t_{\rm conv}^{-1} = v_{\rm w}/l$, where $v_{\rm w}$ is the wind velocity. The maximum energy for electrons is obtained equating the cooling rates with the acceleration rate. The result is $\sim$ 2 GeV. It can be seen that electrons rapidly cool by synchrotron mechanism with timescales of less than 1 s. 
 
 \begin{figure}
\includegraphics[angle=0,width=0.98\hsize]{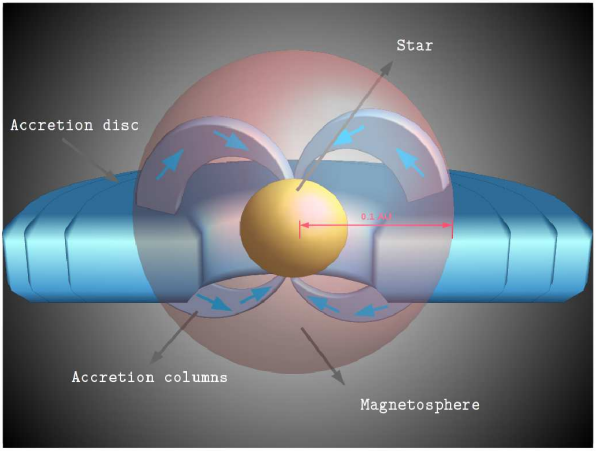}
\caption{Sketch of a T Tauri star adapted from  Feigelson \& Montmerle (1999).}
  \label{fig:magnetosphere}
\end{figure}

The radiative losses for protons are $p\gamma$ interactions with the X-ray radiation field, synchrotron and $pp$ inelastic collisions. The latter losses are the most important for these particles. Figure \ref{fig:perdPro} shows the cooling rates and the acceleration rate for protons. The maximum energy of these particles is much greater than the maximum energy of electrons, $\sim$ 10 TeV. Protons of such an energy satisfy the Hillas criterion for a field of 2$\times$10$^{2}$ G. Acceleration timescales at the higher energies are $\sim10^{5}$ s, larger than the typical durations of flaring X-ray episodes ($\sim10^{4}$ s). Hence, maximum energies of $\sim10$ TeV are more realistic in a steady state approximation to the $\gamma$-ray active period.

\begin{table}
\caption[]{Parameters}
\begin{tabular}{lll}
\hline\noalign{\smallskip}
\multicolumn{2}{l}{Parameter [units]} & values\\[0.001cm]
\hline\noalign{\smallskip}
$l$ &  Flux tube length [cm] & $10^{11}$  \\[0.001cm]
$v_{\rm A}$ & Alfven speed [cm s$^{-1}$] &  6$\times$10$^{7}$ \\[0.001cm]
$\eta$ & Acceleration efficiency & 10$^{-6}$\\[0.001cm]
$a$ & Hadron-to-lepton energy ratio & 100 \\[0.001cm]
$q_{\rm rel}$ & Fractional content of relativistic particles & 10$^{-1}$   \\[0.001cm]
$\alpha$ &Particle injection index & 2.2\\[0.001cm]
$v_{\rm w}$ & Wind velocity [cm s$^{-1}$] &  2$\times10^{7}$  \\[0.001cm]
$B$ &  Magnetic field [G] & 2$\times10^{2}$  \\[0.001cm]
$n$ & Particle density [cm$^{-3}$] &   5$\times$ 10$^{11}$ \\[0.001cm]
\hline\\[0.05cm]
\end{tabular}	
  \label{table}
\end{table}

\subsection{Particle distribution}
We assume an injection function $Q(E)$ that is a power-law, of index 2.2, in the energy of the particles (Drury 1983),
\begin{equation}
 Q(E) = Q_{0} E^{-\alpha}e^{-E/E_{\rm max}}. 
\end{equation}
 The normalization constant $Q_{0}$ for each kind of particle  is obtained from, 
\begin{equation}
 L_{\rm e} = \frac{L_{\rm rel}}{1+a} = V \int_{E_{\rm min}}^{E_{\rm max}} 
Q_{0} E^{-\alpha} E {\rm d} E,
\end{equation}
and
\begin{equation}
 L_{p} = \frac{L_{\rm rel}}{1+1/a} = V \int_{E_{\rm min}}^{E_{\rm max}} 
Q_{0} E^{-\alpha} E {\rm d} E.
\end{equation}
Here $a$ is the hadron-to-lepton energy density ratio. We consider $a = 100$ as observed in cosmic rays in the solar neighborhood (Ginzburg \& Syrovatskii 1964), $L_{\rm rel}$ is the power in the form of relativistic particles and $V$ is the acceleration volume, i.e. the volume of the flux tube. The steady state particle distributions $N(E)$ are obtained solving the transport equation  
in the homogeneous approximation,
\begin{equation}
 \frac{\partial}{\partial E}\biggl[\frac{{\rm d}E}{{\rm d}t}{\bigg\arrowvert}_{\rm loss}N(E)\biggr]+\frac{N(E)}{t_{\rm esc}} = Q(E),
\end{equation}
where $t_{\rm esc} = t_{\rm conv}$. This equation has an exact analytical solution (see Ginzburg \& Syrovatskii 1964): 
\begin{eqnarray}
N(E)= &\biggl\arrowvert \frac{{\rm d}E}{{\rm d}t}\biggl{\arrowvert}_{\rm loss}^{-1}\int_{E}^{E^{\rm max}}{\rm d}E'& Q(E')\nonumber\\ 
& &\times{\exp}\biggl(-\frac{\tau(E,E')}{t_{\rm esc}}\biggr),
\end{eqnarray}
with
\begin{equation}
\tau(E,E')= \int_{E}^{E'} {\rm d}E'' \biggl\arrowvert \frac{{\rm d}E''}{{\rm d}t}\biggr{\arrowvert}_{\rm loss}^{-1}. 
\end{equation}

\begin{figure}[!h]
\includegraphics[angle=270, scale=0.4]{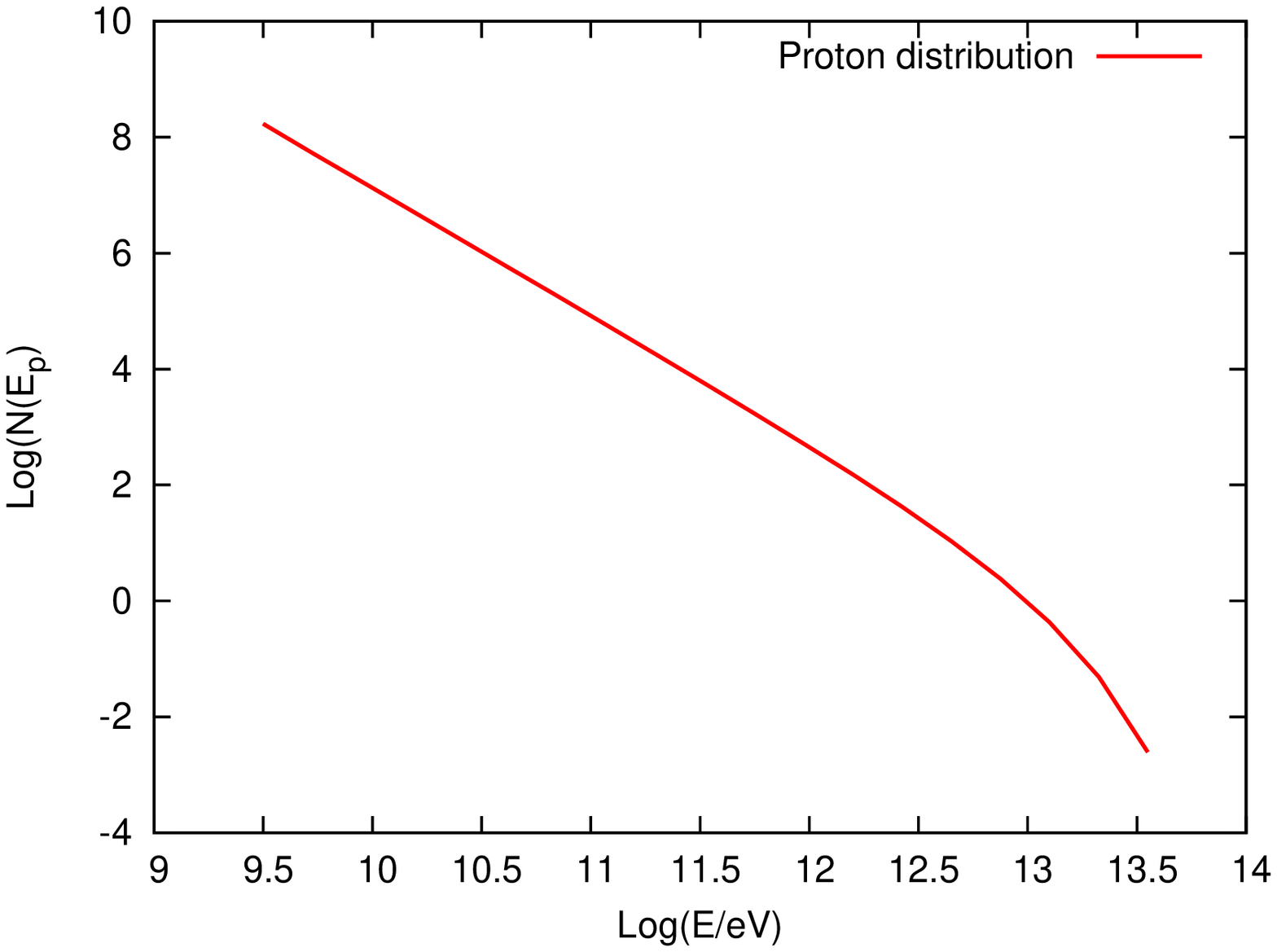}

\caption{Steady state particle distribution for protons.}
\label{Ele}
\includegraphics[angle=270, scale=0.4]{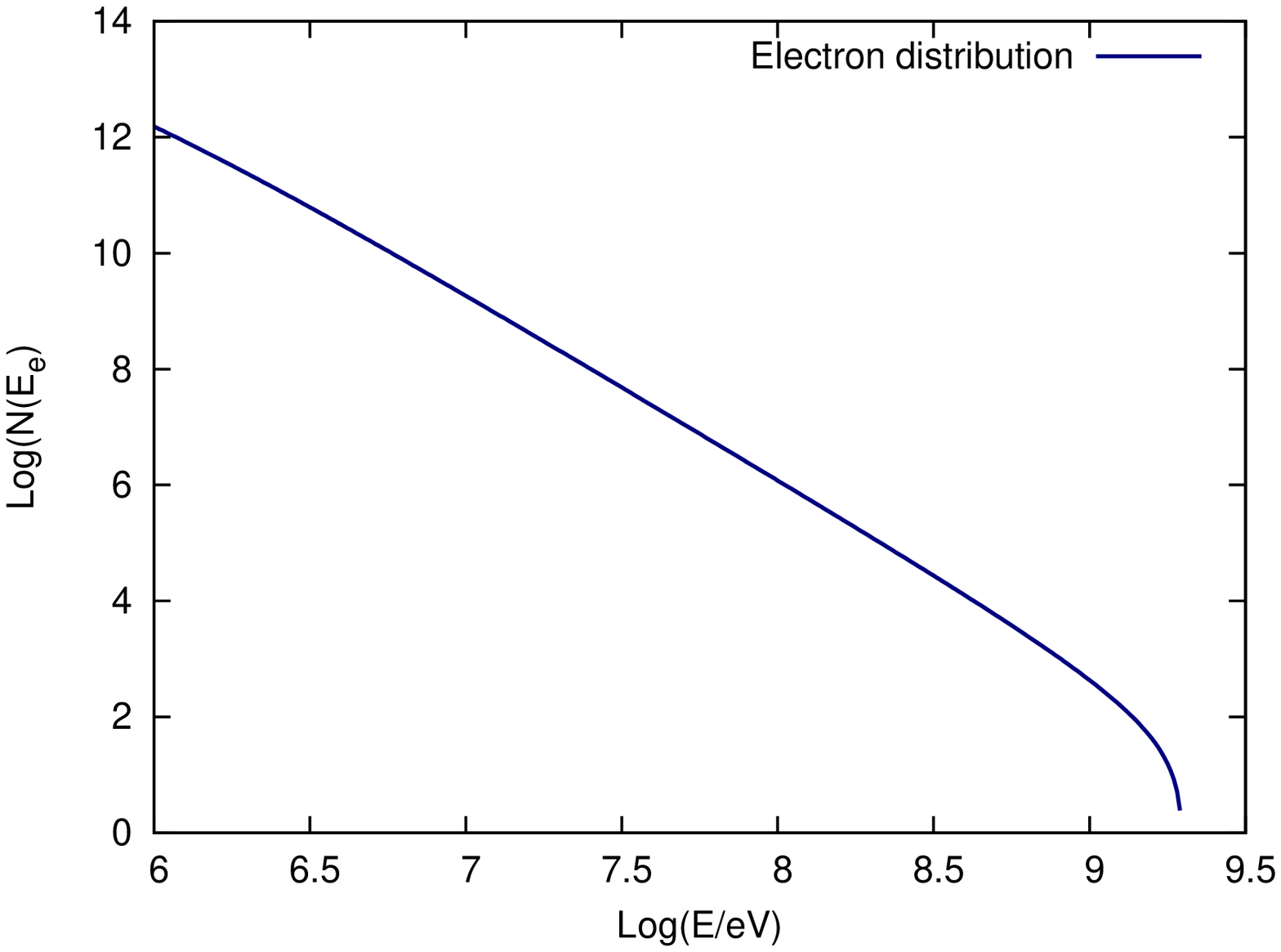}

\caption{Steady state particle distribution for electrons.}
  \label{Pro}
\end{figure}

Figures \ref{Pro} and \ref{Ele} show the steady state particle distribution for protons and electrons. These distributions are valid only as long as we consider intervals much shorter than the timescale of the flare. 

\subsection{Luminosity} 

In order to estimate the luminosity due to electrons we computed synchrotron, relativistic Bremsstrahlung, synchrotron self-Compton (SSC) radiation, and IC up scattering of external seed photons (X-ray and IR radiation fields). We calculate for protons synchrotron and $\gamma$-ray emission from $\pi^{0}$ decay in $pp$ inelastic collisions. We consider also the {synchrotron} contribution from secondary pairs injected by charged pion decay (e.g. Orellana et al. 2007).
The maximum particle density is $\sim$  5$\times$10$^{11}$ cm$^{-3}$. It is localized in the accretion columns. We consider a cylindrical portion of the accretion column of radius $r_{\rm AC}$ $\sim$ $10^{10}$ cm and height $\sim$ 0.1 $r_{\rm AC}$ (e.g. Orlando et al. 2010). This volume is where most of the relativistic interactions with matter take place\footnote{The accretion disk is truncated at the edge of the magnetosphere. The exterior disk material has no major impact in the calculations since the interacting matter is in the accreting columns that penetrate the magnetosphere.}. For more details on high-energy processes see Vila \& Aharonian (2009) and Vila (2010).

\subsubsection{Internal absorption}

Attenuation of the radiation produced by photon annihilation is expected in T Tauri stars. The opacity produced by a photon field with density $n_{\rm ph}(\epsilon)$ and photon energy $\epsilon$ is
\begin{eqnarray}
\tau (E_{\gamma}) = &&
  \frac{1}{2}\int_{l} \int_{\epsilon_{\rm th}}^{\epsilon_{\rm max}} \int_{-1}^{u_{\rm max}} (1-u) \, \sigma_{\gamma\gamma}(E_{\gamma}, \epsilon,u)
n_{\rm ph}(\epsilon)\nonumber\\
&&\times {\rm d}u {\rm d}\epsilon {\rm d}l .
\end{eqnarray}
Here $u = \cos \vartheta$, $\vartheta$ is the angle between the momenta of the colliding photons, $l$ is the photon path, and $\sigma_{\gamma\gamma}(E_{\gamma}, \epsilon,u)$ is the cross section for photon annihilation (Gould \& Schr\'eder 1967).

\begin{figure}[!h]
\resizebox{\hsize}{!}{\includegraphics[angle=270]{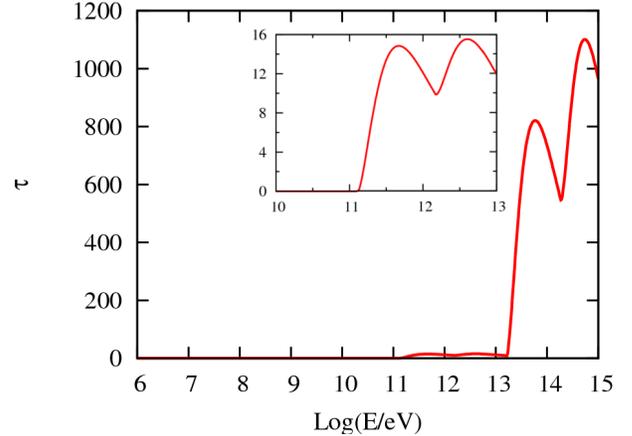}}
\caption{Opacity curve as a function of energy $E$.}
\label{Opa}
\end{figure}

The absorbing photon fields are those generated within the system and the strong blackbody radiation field from the surroundings: the disk (IR), the star, and the X-rays from the accreting plasma.To estimate  the disk infrared photon field we adopt an internal radius for the disk of $R_{\rm D}$ $\sim$ 120 AU (see Fig. \ref{fig:magnetosphere}), and a temperature of $T$ $\sim$ 30 K (Dutrey et al. 1994). The X-ray temperature is taken to be $T$ $\sim$ 10$^{7}$ K (e.g. Feigelson \& Montmerle 1999). Regarding the star, we consider a temperature of $T_{\star}$ $\sim$ 4$\times$10$^{3}$ K, and a radius $R_{\star}$ $\sim$ 2 $R_{\odot}$. The opacity $\tau$ (attenuation $e^{-\tau}$) as a function of $E_{\gamma}$ is shown in Fig. \ref{Opa}. The absorption is almost complete above 100 GeV.

\section{Application to possible $\gamma$-ray emitting T Tauri stars}

\begin{figure}
\includegraphics[angle=0, scale=0.28]{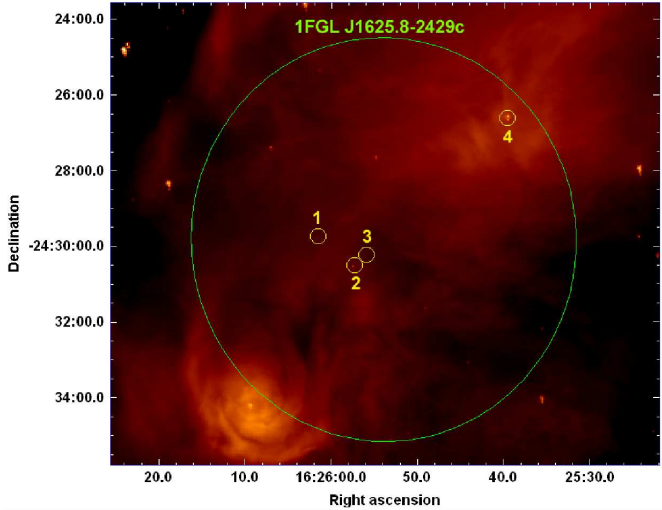}
\caption{GLIMPSE infrared image in the $8.0$ $\mu$m band showing the contents of the
\bicho\ error circle towards the $\rho$ Ophiuchi cloud. Several T Tauri stars are
consistent with the {\it Fermi} $\gamma$-ray source position. They are labelled from 1 to 4 in decreasing
order or right ascension. Axis coordinates are of equatorial J2000.0-type.}
 \label{ir_map}
\end{figure}

\begin{figure}
\includegraphics[angle=0, scale=0.28]{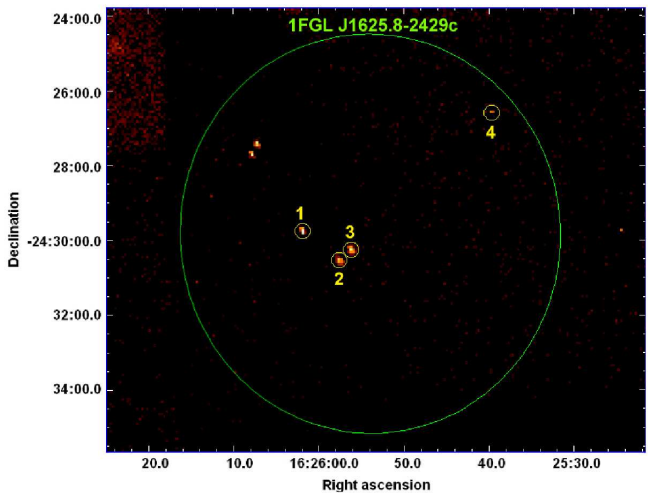}
\caption{Composite X-ray image of the \bicho\ error circle obtained with 
the {\it Chandra} ACIS camera in the energy range 
0.1-10 keV (Dataset identifier: ADS/Sa.CXO\#obs/00618).  Numbers indicate the T Tauri stars consistent with this {\it Fermi} 
source in decreasing order of right ascension. All of these stars are X-ray emitters.}
 \label{x_map}
\end{figure}

\begin{figure}
\includegraphics[angle=0, scale=0.28]{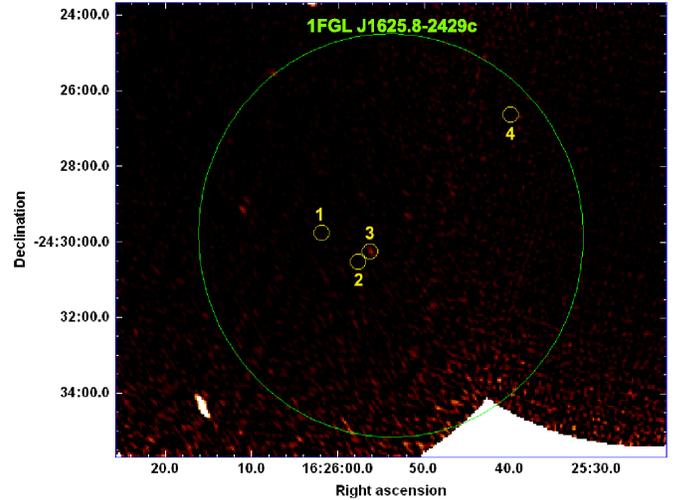}
\caption{Radio mosaic of the \bicho\ error circle as observed with the VLA at the 6 cm wavelength. Numbers indicate
the T Tauri stars consistent with this {\it Fermi} source in decreasing order of right ascension. Only the T Tauri star \# 3
is detected in the radio.}
 \label{radio_map}
\end{figure}

After performing several catalog cross-correlations, a new possible association with nearby T Tauri stars inside the well-known $\rho$ Ophiuchi star forming region clearly emerges for the  {\it Fermi} source 1FGL J1625.8 $-$2429c. Indeed, inside the $95\%$ confidence error ellipse of this {\it Fermi} source we find four T Tauri stars: 
2MASS J16260160$-$2429449 (Casanova et al. 1995),
2MASS J16253958$-$2426349 (Wilking et al. 1989),
JCMTSF J162556.8$-$243014 (Di Francesco et al. 2008),
and 2MASS J16255752$-$2430317 (Grasdalen et al. 1973).
Their observational properties are summarized in Table \ref{tableIII}.
In Figs. \ref{ir_map}, \ref{x_map}
and \ref{radio_map} we show this field ($l=$353\grp 0, $b=$17\grp0)

as observed in the infrared, X-ray and radio wavelengths. 
These images have been produced using public data retrieved (and calibrated when necessary)
from the {\it Spitzer}-GLIMPSE, {\it Chandra} and NRAO Very Large Array (VLA) archives, respectively.
As discussed below, we tentatively suggest that this {\it Fermi} source 
might be the result of the emission of at least these four Tauri stars
that lay inside the location error box of 1FGL J1625.8 $-$2429c.

\begin{table}
\caption[]{{\it Fermi} integrated fluxes and calculated fluxes.}
\begin{tabular}{lll}
\hline\noalign{\smallskip}
\multicolumn{1}{l}{l}{Energy range} & Flux $\pm$ error [cm$^{-2}$ s$^{-1}$] & Computed flux \\[0.001cm]
\hline\noalign{\smallskip}
100 MeV-300 MeV  & 1.9$\times 10^{-7}$ $\pm$ $1.0\times 10^{-8}$ &  6.30$\times 10^{-8}$ \\[0.001cm]
300 MeV-1 GeV & 2.2$\times 10^{-8}$ $\pm$ 0.6$\times 10^{-8}$ &   1.89$\times 10^{-8}$\\[0.001cm]
1 GeV-3 GeV  & 4$\times 10^{-9}$ $\pm$ 0.7$\times 10^{-9}$& 4.82$\times 10^{-9}$\\[0.001cm]
3 GeV-10 GeV  & 8.5$\times 10^{-10}$ $\pm$ 2.2$\times 10^{-10}$& 1.21$\times 10^{-9}$ \\[0.001cm]
10 GeV-100 GeV  & 2.1$\times 10^{-10}$ $\pm$ $1.0\times 10^{-11}$& 3.76$\times 10^{-10}$ \\[0.001cm]
\hline\\[0.05cm]
\end{tabular}	
\label{tableII}
\end{table}

\begin{table*}
\caption[]{Properties of T Tauri stars proposed as \bicho\ counterpart.}
\begin{tabular}{cccccc}
\hline\noalign{\smallskip}
\multicolumn{1}{c}{Star \#} & Name      & Mag. & X-ray counterpart & 0.5-7 keV flux            & 6-cm flux density   \\[0.001cm]
                            &           &           &  CXO                 &  (erg s$^{-1}$ cm$^{-2}$) &  (mJy)                    \\[0.001cm]
\hline\noalign{\smallskip}
1            & 2MASS J16260160$-$2429449  &  $R=17.2$  &  J162601.6$-$242945  &  $2.30 \times 10^{-13}$    &  $-$              \\
2            & 2MASS J16255752$-$2430317  &  $V=16.4$  &  J162557.5$-$243031  &  $4.52 \times 10^{-13}$    &  $-$              \\
3            & JCMTSF J162556.8$-$243014/ &  $R=16.8$  &  J162556.0$-$243014  &  $4.95 \times 10^{-13}$    &  $0.30 \pm 0.03 $  \\
             & 2MASS J16255609$-$2430148  &            &                          &                             &                   \\
4            & 2MASS J16253958$-$2426349  &  $V=19.1$  &  J162539.5$-$242634  &  $4.90 \times 10^{-14}$    &  $-$              \\
\hline\\[0.05cm]
\end{tabular}
\label{tableIII}
\end{table*}

In order to estimate the {\it a priori} probability of a pure chance association we have implemented Monte Carlo simulations of computer-generated {\it Fermi} sources following the approach developed by Romero et al. (1999) for unidentified EGRET sources. After $10^4$ simulations of artificial {\it Fermi} populations we find 47 coincidences at 1-degree binning and 4 at 2-degree binning , indicating a probability of chance association of  $\sim 10^{-3}$. These results do not change with larger samples (we run up to $10^6$ simulations).

\begin{figure*}[]
\centering
\resizebox{\hsize}{!}{\includegraphics[angle=270]{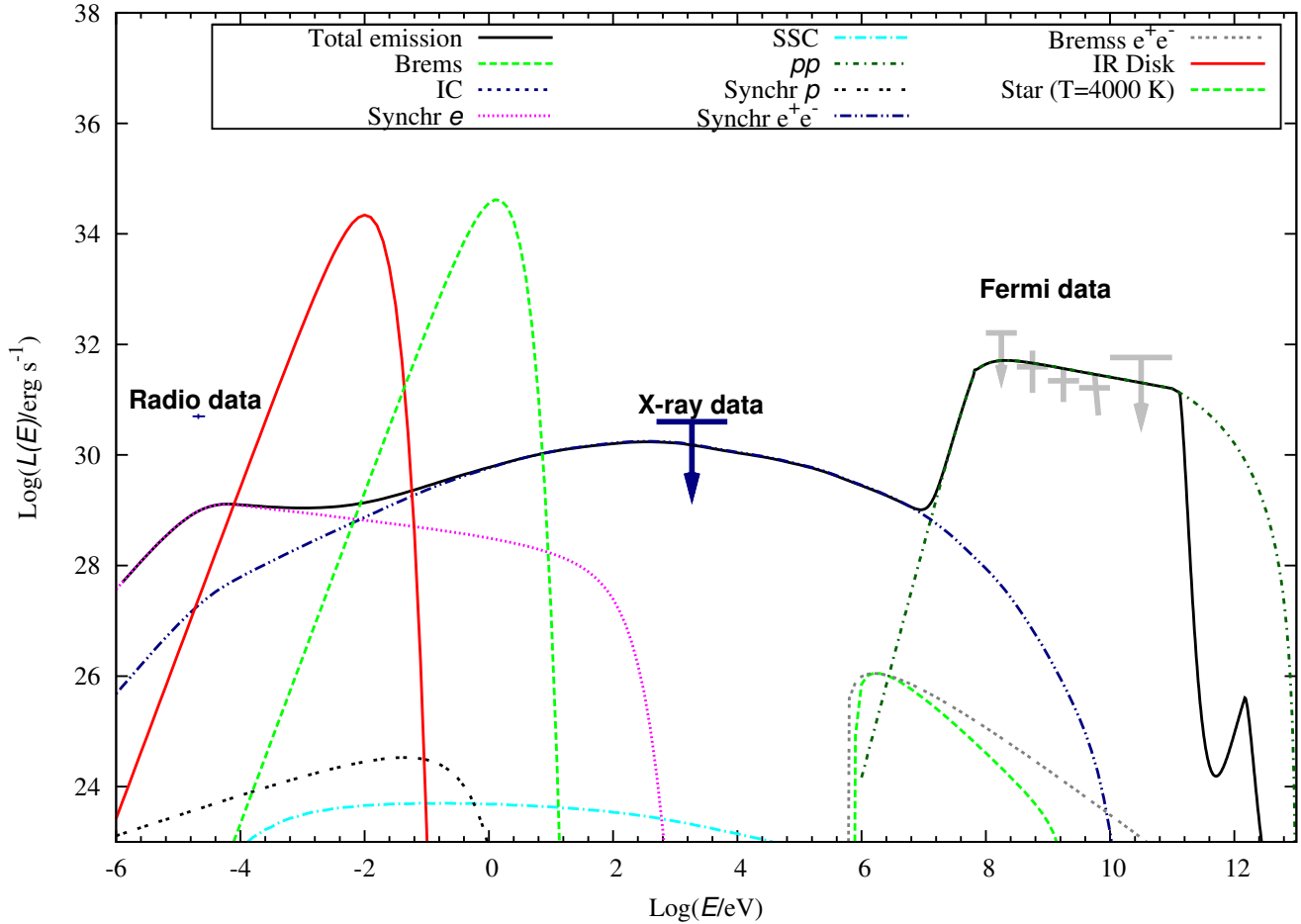}}
\caption{Computed non-thermal luminosity and {\it Fermi} upper bounds for the four  T Tauri stars, assuming 
a distance of 120 pc. The spectral energy distribution is corrected by photon absorption. Model parameters as in Table 1.}
  \label{LumiF}
\end{figure*}

Let $F(E_{1},E_{2}) \pm \bigtriangleup[F(E_{1},E_{2})]$ be the integrated {\it Fermi} flux in the energy range [$E_{1}, E_{2}$] and its error.
To reproduce the observed fluxes, we consider in first approximation that 
the four T Tauri stars emit the same $\gamma$-ray luminosity. Then, we can compute the integrated flux from: 
\begin{equation}
F=4\int_{E_{1}}^{E_{2}} \frac{ L_{\gamma}(E)}{4{\pi}d^{2} E^{2}} {\rm d}E .
\end{equation}
Here $L_{\gamma}(E)$ is the total $\gamma$-ray luminosity produced by an individual T Tauri star according to our model ($E >$ 20 MeV) and $d$ is the distance to $\rho$ Ophiuchi cloud, $\sim$ 120 pc (Loinard et al. 2008).  The integrated {\it Fermi} fluxes (Abdo et al. 2010) and the calculated fluxes in  five energy bands are shown in Table \ref{tableII}.
Figure \ref{LumiF} shows the luminosity obtained with the model and the upper bound given by {\it Chandra} and {\it Fermi} data.  Radio data from VLA are also shown, but this radiation is surely a combination of non-thermal and thermal emission and hence must be considered just as an upper value to constrain the model.

\section{Discussion}

We see that for a reasonable set of physical parameters of T Tauri stars, a relatively weak $\gamma$-ray source can be produced. The intrinsic luminosity is not high ($L_{\gamma}$ $\sim$ 10$^{31}$ erg s$^{-1}$ at $\sim 1$ GeV), but since the stars are very nearby the flux is significant. If such an association is confirmed, the T Tauri stars in $\rho$ Ophiuchi would be the nearest $\gamma$-ray sources to the solar system detected so far. A caveat is in order, however, considering the potential effects of the galactic cosmic rays with the molecular material in $\rho$ Ophiuchi. This emission has been subtracted from the {\it Fermi} $\gamma$-ray data (Abdo et al. 2010), but there could be more matter than detected so far. 

In the case that matter is underestimated in $\rho$ Ophiuchi, we calculate the required density excess to explain the origin of the high-energy emission from the source \bicho. With this aim we assume that the whole $\gamma$-ray source is the result of passive ``illumination'' of cloud material by cosmic rays (e.g. Aharonian \& Atoyan 1996).
We adopt a standard cosmic-ray spectrum in $\rho$ Ophiuchi. Proton-proton inelastic collisions with the ambient medium are the main radiative process contributing to the high-energy electromagnetic spectrum. The cosmic-ray proton spectrum has a  power law of index $\alpha$ = 2.7 and  the enhancement parameter is taken to be 1 (i.e. absence of local acceleration). The normalization constant  is obtained from the energy density of cosmic rays in the solar neighborhood, i.e. ${\omega}_{\rm CR}$ $\sim$ 1 eV cm$^{-3}$ (e.g. Ginzburg \& Syrovatskii 1964). With these figures and the {\it Fermi} flux we obtain an average density  of $\left\langle n \right\rangle $ $\sim$ 10$^{4}$ cm$^{-3}$, one order of magnitude greater than current estimates\footnote{We remark that the cosmic-ray energy density in $\rho$ Ophiuchi is not known with any accuracy, a fact that has impact in the estimates. However, a difference of one order of magnitude would be, at least, peculiar.} (e.g. Crutcher 1991).

If the effects of the galactic cosmic rays with the interstellar  material are responsible for the {\it Fermi} source, then {\bicho} would be the nearest passive $\gamma$-ray source detected  outside the solar system.

\section{Conclusions}

We have found that under some assumptions T Tauri stars might be  responsible for some nearby  {\it Fermi} sources.  We have presented  a simplified model for the high-energy emission of this type of stars, that agrees with the available multiwavelength observations. T Tauri stars might be a new  class of galactic $\gamma$-ray  sources in the Galatic plane. Based on this new scenario, 1FGL J1625.8-2429c is the first  candidate for collective
$\gamma$-ray emission from low-mass protostars. If the association with
the $\rho$ Ophiuchi cloud is confirmed, 
it would  be the closest $\gamma$-ray source to the Solar System.  
This statement still holds even in the alternative case where 
the detected $\gamma$-rays were simply due to cosmic rays 
interacting with the cloud ambient gas.

More complex  models can be developed in the future, using new and deeper observations. The energy where the cutoff lays might be established in the future by new  Cherenkov  instruments like CTA.

\acknowledgments
We thank the referee Elisabete M. de Gouveia Dal Pino for constructive and insightful suggestions.

The authors acknowledge support by grant AYA2010-21782-C03-03 from the Spanish government,
and FEDER funds. This work has been also supported 
by the Consejer\'{\i}a de Innovaci\'on, Ciencia y Empresa (CICE)
of Junta de Andaluc\'{\i}a as research group FQM-322 and excellence fund FQM-5418.
This research has made use of the SIMBAD database, operated at CDS, Strasbourg, France. 
-the NRAO is a facility of the NSF operated under cooperative agreement by Associated Universities, Inc-. This research also has made use of data from {\it Chandra X-ray Observatory} Center, which is operated by the Smithsonian Astrophysical Observatory for and on behalf of the National Aeronautics Space Administration; and of the Spitzer-GLIMPSE database, which is operated by the Jet Propulsion Laboratory, California Institute of Technology. 
G.E.R. and M.V. del V. are supported by the Argentine Agency CONICET (PIP 0078) and ANPCyT (PICT 2007-00848).

\end{document}